\documentstyle[12pt,epsfig,axodraw,a4]{article} 
\newcommand{\vs}{\vspace{-0.25cm}}
\begin{document} 
\begin{center}
\large{\bf Note on spin-orbit interactions in nuclei and
  hypernuclei}\footnote{Work supported in part by BMBF, GSI and by the DFG
  cluster of excellence: Origin and Structure of the Universe.}

\bigskip

N. Kaiser and W. Weise\\

\medskip

{\small Physik Department, Technische Universit\"{a}t M\"{u}nchen,
    D-85747 Garching, Germany}

\end{center}

\medskip

\begin{abstract}
A detailed comparison is made between the spin-orbit interactions in $\Lambda$ 
hypernuclei and ordinary nuclei. We argue that there are three major
contributions to the spin-orbit interaction: 1) a short-range component 
involving scalar and vector mean fields; 2) a ''wrong-sign'' spin-orbit term 
generated by the pion exchange tensor force in second order; and 3) a
three-body term induced by two-pion exchange with excitation of virtual 
$\Delta(1232)$-isobars (a la Fujita-Miyazawa). For nucleons in nuclei the 
long-range pieces related to the pion-exchange dynamics tend to cancel, leaving 
room dominantly for spin-orbit mechanisms of short-range origin (parametrized
e.g. in terms  of relativistic scalar and vector mean fields terms). In  
contrast, the absence of an analogous $2\pi$-exchange three-body contribution
for $\Lambda$ hyperons in hypernuclei leads to an almost complete cancellation
between the short-range (relativistic mean-field) component and the  
''wrong-sign'' spin-orbit interaction generated by second order $\pi$-exchange
with an intermediate $\Sigma$ hyperon. These different balancing mechanisms  
between short- and long-range components are able to explain simultaneously
the  very strong spin-orbit interaction in ordinary nuclei and the remarkably 
weak spin-orbit splitting in $\Lambda$ hypernuclei.
\end{abstract}

\bigskip

PACS: 21.30.Fe, 21.80+a, 24.10.Cn\\

\section{Introduction}
The microscopic understanding of the dynamical origin behind the strong nuclear 
spin-orbit force is still one of the key questions in nuclear physics. The 
analogy with the spin-orbit interaction in atomic physics gave the hint that it
could be a relativistic effect. This idea has lead to the phenomenologically
successful scalar-vector mean field models for nuclear structure calculations 
\cite{walecka,ringreview}. In these models the nucleus is described as an 
ensemble of independent Dirac quasi-particles moving in self-consistently 
generated scalar and vector mean fields. With nucleon Fermi momenta typically 
less than a third of the (free) nucleon mass, the motion of the nucleons in
nuclei is non-relativistic on average. Signatures of relativity are 
nonetheless manifest in the large spin-orbit coupling which emerges in that
framework naturally from the interplay of the individually large scalar and
vector mean fields of opposite signs. Their sum balances such as to produce a
relatively weak central potential, whereas their difference coherently generates
the strong spin-orbit potential \cite{duerr,miller,BW}. 

In the context of QCD sum rule calculations these scalar and vector 
mean fields can be related to the leading changes of the scalar quark
condensate $\langle \bar qq \rangle$ and the quark density $\langle q^\dagger
q \rangle$ at finite baryon density. This connection has been utilized in
Ref.\cite{finelli} to derive a relativistic nuclear energy density
functional constrained by low-energy QCD. In such an approach nuclear binding 
originates primarily 
through pionic fluctuations (i.e. two-pion exchange calculated with in-medium
chiral perturbation theory) while the spin-orbit interaction results
from the strong scalar and vector mean fields related to changes of the 
condensate structure of the QCD vacuum at finite baryon density. Nuclear 
structure calculations of spherical and deformed nuclei performed within 
this approach reach the same level of accuracy, in comparison with data, as
the best phenomenological relativistic mean field models.    

In a phenomenological boson exchange picture the relativistic spin-orbit 
interaction is modeled by the exchange of a scalar boson ($``\sigma$'') and a
 vector boson ($``\omega$'') between nucleons, simulating short-range dynamics 
but ignoring the physics at intermediate and long ranges characteristic of the 
average distance scale between nucleons in  nuclei. Of particular interest in 
the context of the spin-orbit interaction are the effects
from the exchange of two pions. As
a matter of fact, the dominant tensor interaction from one-pion exchange
has a spin- and momentum dependence that produces, in second order, the
spin-orbit coupling $i(\vec \sigma_1+ \vec\sigma_2)\cdot(\vec q \times \vec 
p\,)$ in momentum  space \cite{nnpap}. Here, $\vec \sigma_{1,2}$ are the usual
Pauli spin-operators of the nucleons; $\vec p$ and $\vec q$ denote the
center-of-mass momentum  
and the momentum transfer in the elastic $NN$ scattering process. The explicit 
calculation of the relevant pion loop integral representing iterated one-pion 
exchange (the second-order tensor force) gives the following isoscalar
spin-orbit $NN$ scattering amplitude at threshold ($\vec p = \vec q = 0)$: 
\begin{eqnarray}
V_{so}^{(\pi\pi)} = -{g_A^4M_N\over 64\pi f_\pi^4 \,m_\pi} <0~,\nonumber 
\end{eqnarray}
with the nucleon axial vector constant $g_A \simeq 1.3$ and the pion decay
constant $f_\pi = 92.4$ MeV. At first sight, this contribution has the
''wrong'' sign in comparison to the one from vector boson exchange,
\begin{eqnarray}
 V_{so}^{(\omega)}={3g_{\omega  N}^2\over4 M_N^2\,m_\omega^2}>0~,\nonumber
 \end{eqnarray}
 or scalar boson exchange,
\begin{eqnarray}
 V_{so}^{(\sigma)}={g_{\sigma  N}^2\over 4M_N^2\,m_\sigma^2}>0~~.\nonumber
 \end{eqnarray}
As it stands the ``wrong-sign'' spin-orbit interaction from iterated 
$1\pi$-exchange in the two-nucleon system must also show up  
in nuclear many-body systems.
 
On the other hand it has long been known that calculations based on Hamiltonians
with realistic two-nucleon potentials (fitting accurately all empirical
$NN$ phase shifts and mixing angles) cannot correctly predict the observed 
spin-orbit splittings of nuclear levels in light nuclei \cite{pand,pieper}. A 
suitably chosen three-body force was then implemented  to account for this
deficiency. In fact, one of the basic motivations for the genuine three-nucleon 
interaction introduced originally by Fujita and Miyazawa \cite{fujita} was the
study  of such spin-orbit splittings. The 
Fujita-Miyazawa three-nucleon interaction involves the exchange of two pions 
between the three nucleons, where the central nucleon which couples to both 
pions is excited to a $\Delta(1232)$ resonance in the intermediate state. When 
considered in the vacuum this $2\pi$-exchange process with excitation of 
virtual $\Delta(1232)$ isobars generates the long-range tail of the isoscalar 
central attraction between nucleons. However, in the medium the Pauli blocking 
of intermediate nucleon states below the Fermi surface changes the (orbital) 
angular momentum balance and an additional spin-orbit interaction (which
increases  with density) emerges. This long ranged pion-induced three-body 
interaction provides an additional important contribution to the spin-orbit
force in nuclear many-body systems.  

\newpage
In summarizing this introductory discussion, we thus identify three major 
sources of the spin-orbit interaction in nuclear systems:
\begin{itemize}
\item{\bf Short distance dynamics of coherently acting scalar and vector fields}
\end{itemize}
These effects are frequently parametrized in terms of sigma-omega exchange 
models. Since only the ratios $g^2/m^2$ of squared coupling constants and
boson masses appear in any relevant calculated quantity, an equivalent
effective field theory approach with $NN$ contact interactions, encoding short
distance dynamics with coupling strength parameters $G_V = g_{\omega N}^2/m_\omega^2$ 
and $G_S = g_{\sigma N}^2/m_\sigma^2$, will yield identical output. In alternative
QCD sum rule descriptions these terms can be thought of as originating from 
leading in-medium changes of the quark condensate $\langle\bar{q}q\rangle$ and
the quark density $\langle q^\dagger q\rangle$.
\begin{itemize}
\item{\bf Intermediate and long range spin-orbit dynamics}
\end{itemize}
induced by the pion exchange tensor force in second order, with Pauli blocking 
acting on intermediate nucleon states. This mechanism has an analogous
counterpart in hypernuclei where the $\Lambda N$ interaction involves an
intermediate $\Sigma$ hyperon between pion exchanges. 
\begin{itemize}
\item{\bf Three-body spin-orbit interaction}
\end{itemize}
of the Fujita-Miyazawa type, generated by two-pion exchange with intermediate 
excitation of a virtual $\Delta$ isobar. It is this latter important
ingredient that has no counterpart in $\Lambda$ hypernuclei. As a consequence,
the balance of all three major contributions to the spin-orbit is
qualitatively different between nuclei and hypernuclei, as we shall demonstrate.
  
\section{Spin-orbit coupling for nucleons}          
We begin with a reminder of facts about nuclear spin-orbit interactions.
In order to quantify the spin-orbit terms discussed before it is
convenient to consider the energy density functional which serves as a
general starting point for nuclear structure calculations (of heavy nuclei) 
within the self-consistent mean-field approximation \cite{reinhard}. The 
spin-orbit coupling term in the nuclear energy density functional has the 
form: 
\begin{equation} {\cal E}_{\rm so}[\rho,\vec J\,] = F_{\rm so}(\rho)\,\vec  \nabla  
\rho\cdot\vec J\,. \end{equation}
It is constructed from the gradient of the (ordinary) nuclear density 
distribution $\rho(\vec r\,)$ and the so-called spin-orbit density, defined by:
\begin{equation} \vec J(\vec r\,)=\sum_{\alpha \in \rm    occ}\Psi_\alpha^\dagger(\vec r\,)
i\, \vec \sigma \times \vec \nabla\Psi_\alpha( \vec r\,) \,. \end{equation}
The strength of the nuclear spin-orbit coupling is measured by the density 
dependent strength function $F_{so}(\rho)$.\footnote{The frequently used
single-particle spin-orbit potential strength is $U_{Nls}(\rho)=2\rho F_{so}
(\rho)$.} In the non-relativistic Skyrme phenomenology the latter is treated
as an adjustable constant parameter with a typical value of $F_{so}(\rho)
\equiv 3W_0/4 \simeq 90\,$MeV\,fm$^5$. It is worth noting that this value of
the spin-orbit coupling strength is quite stable within the huge class of
effective Skyrme forces considered in the literature \cite{reinhard,skyrme}. 

The basic framework for calculating the nuclear spin-orbit energy density 
functional ${\cal E}_{\rm so}[\rho,\vec J\,]$ is the density matrix expansion of 
Negele and Vautherin \cite{negele}. It generalizes the step-function  
distribution $\theta(k_f -|\vec p\,|)$ of nucleon momentum states for
infinite nuclear matter to the situation of an inhomogeneous many-nucleon 
system with a local density $\rho(\vec r\,)$ and a local spin-orbit density 
$\vec J(\vec r\,)$. Using this technique the Hartree diagrams of scalar
and vector boson exchange between nucleons lead to the well-known expression
for the spin-orbit coupling strength:  
\begin{equation} F_{so}(\rho)^{(\sigma,\omega)} = {1 \over 4M^{*2}(\rho)} \bigg(
{g_{\sigma N}^2 \over m_\sigma^2}+{g_{\omega N}^2\over m_\omega^2}\bigg)=
{G_S+G_V\over 4M^{*2}(\rho)}~.
\label{sigomega}
\end{equation}  
Choosing typical values for the ratios of coupling constants to boson masses, 
$G_S \simeq G_V \simeq 11\,$fm$^2$,
together with an effective (in-medium) nucleon mass of $M^*(\rho_0) \simeq 0.73 
M_N$ one reproduces the empirical value of $90\,$MeV\,fm$^5$ for the spin-orbit 
coupling strength. 

\begin{figure}
\begin{center}
\includegraphics[scale=0.8,clip]{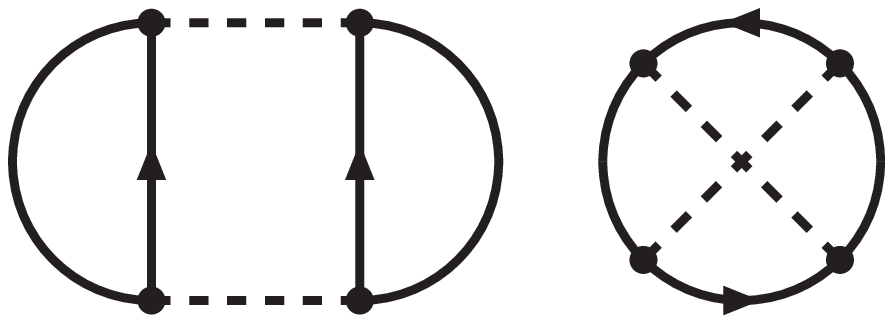}
\end{center}
\vspace{-0.2cm}
{\it Fig.\,1: Three-loop Hartree and Fock related to iterated one-pion
 exchange.}
\end{figure}

Let us now turn to the ''wrong-sign'' spin-orbit interaction from the iterated
one-pion exchange tensor force. The corresponding three-loop Hartree and Fock 
diagrams are shown in Fig.\,1. Their contributions to the density dependent
strength function $F_{so}(k_f)$ have been evaluated analytically in
Ref.\cite{efun}. For the purpose of illustration we reproduce here the 
expression for the dominant two-body Hartree contribution:
\begin{equation} 
F_{\rm so}(\rho)^{(\pi\pi,H2)}={g_A^4 m_\pi M_N\over 64\pi\,f_\pi^4}\Bigg\{ 
{1\over m_\pi^2+4k_f^2}-{3\over 8k_f^2}\ln{m_\pi^2+4k_f^2\over m_\pi^2} 
\Bigg\} \,, 
\label{pipi}
\end{equation}
with the Fermi momentum $k_f$ related to the nucleon density in the usual 
way: $\rho= 2k_f^3/3\pi^2$. When evaluated at nuclear
matter saturation density, $\rho_0 = 0.16\,$fm$^{-3}$, the expression in
Eq.(4) leads to a large ''wrong-sign'' spin-orbit coupling strength of 
$F_{\rm  so}(\rho_0)^{(\pi\pi,H2)} = -86.5\,$MeV\,fm$^5$. Note that this term scales 
linearly with the large nucleon mass $M_N=939$\,MeV. This unfamiliar behavior
of a spin-orbit coupling is obviously not a relativistic effect. It comes from
the energy denominator of the iterated pion-exchange, which is proportional to
the difference of small nucleon kinetic energies. As it stands, this large
spin-orbit Hartree term (\ref{pipi}) would simply cancel the scalar-vector
mean field term of Eq.(\ref{sigomega}) at normal nuclear matter density.

\begin{figure}
\begin{center}
\includegraphics[scale=0.5,clip]{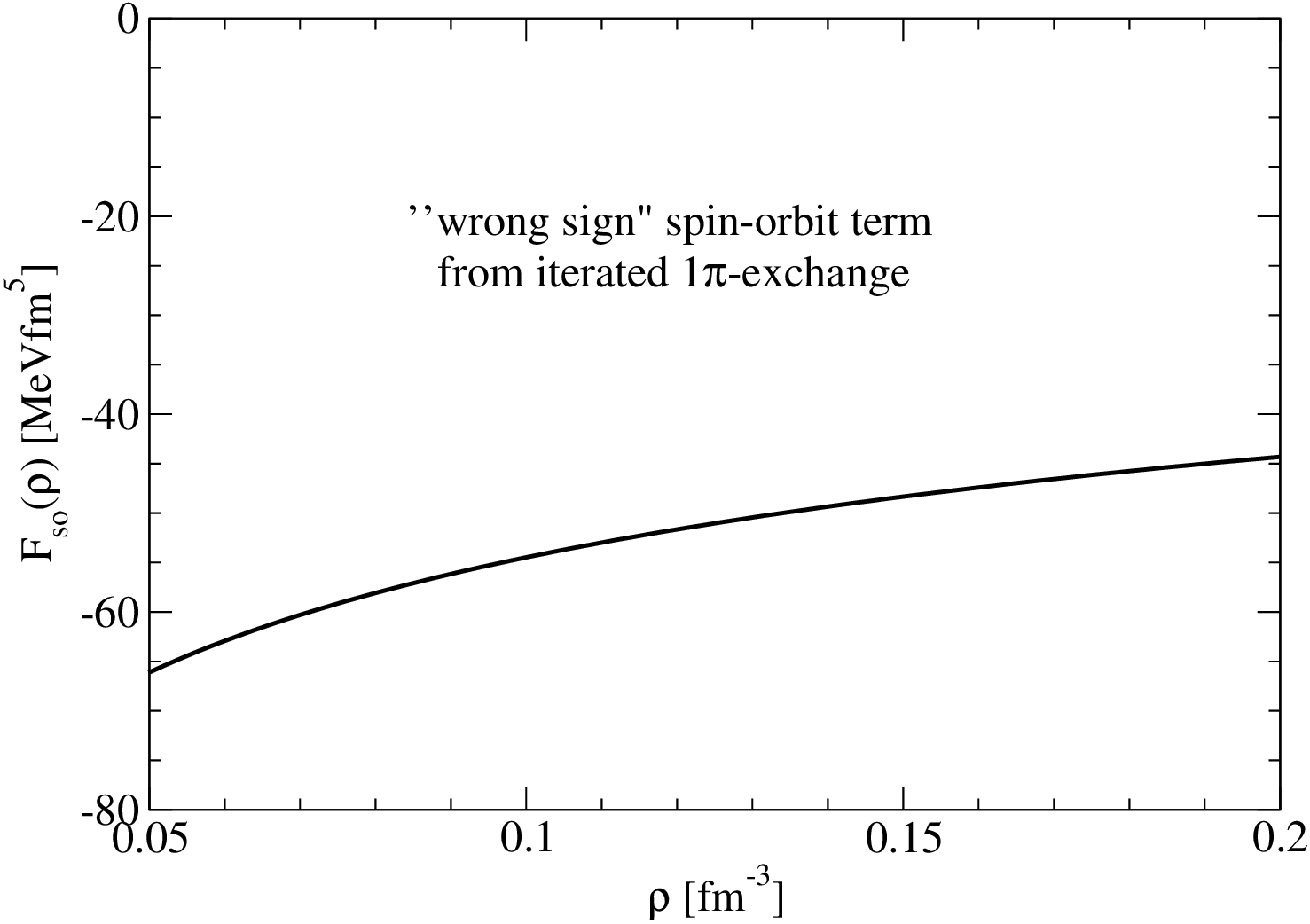}
\end{center}
\vspace{-0.2cm}
{\it Fig.\,2: ''Wrong-sign'' spin-orbit coupling strength generated by the 
second order $1\pi$-exchange tensor force \cite{efun}.}
\end{figure}

The complete spin-orbit coupling strength from iterated
$1\pi$-exchange, with Fock term and Pauli blocking corrections
included \cite{efun}, is shown in Fig. 2 as a function of density. At nuclear 
matter saturation density $\rho_0$ a sizeable spin-orbit coupling strength
$F_{so}(\rho_0)^{(\pi\pi)} =  -47.4\,$MeV\,fm$^5$, amounting to about one half
of the empirical value but with the ''wrong'' sign, still remains.

We demonstrate now that this ''wrong-sign'' spin-orbit term is compensated by
the three-body effects related to $2\pi$-exchange with virtual 
$\Delta(1232)$ isobar excitation. The dominant Hartree diagram in Fig.\,3 leads
to a spin-orbit strength function \cite{3bodyso}: 
\begin{equation} F_{\rm so}(\rho)^{(\Delta,H3)}= {g_A^4 \over 8\pi^2
\Delta f_\pi^4} \Bigg\{ {m_\pi^2k_f+2k_f^3 \over m_\pi^2+4k_f^2}-{m_\pi^2\over 4k_f}
\ln{m_\pi^2+4k_f^2\over m_\pi^2} \Bigg\} 
\,, \end{equation} 
with $\Delta = 293\,$MeV the delta-nucleon mass splitting which, notably, is a
''small" scale just like the pion mass $m_\pi$ and the nuclear Fermi momentum
$k_f$ when compared with the spontaneous chiral symmetry breaking scale,
$4\pi\,f_\pi\simeq 1.2\,$GeV. Again, this contribution to the spin-orbit coupling 
is {\it  not} a relativistic effect. It originates primarily from the spin- and
momentum dependence of the pion-baryon coupling and the fact that all three
participating nucleons are from the filled  
Fermi sea. Adding the very small contribution from the Fock diagrams
\cite{3bodyso}, the spin-orbit coupling strength $F_{\rm so}(\rho)^{(\Delta)}$
shown in Fig.\,4 results. As expected for a genuine three-body effect it grows 
with the nucleon density $\rho$. At nuclear matter saturation density one
reads off the value $F_{\rm so}(\rho_0)^{(\Delta)}=47.5\,$MeV\,fm$^5$ which
cancels the ''wrong-sign'' spin-orbit term from the second order 
$1\pi$-exchange tensor force. Further relativistic spin-orbit effects from the
irreducible two-pion exchange (scaling as $1/M_N$ with the nucleon mass) 
have been investigated in Ref.\cite{realnn}. These come out to be relatively
small such that they do not affect in a significant way the cancellation 
mechanism between ''wrong-sign'' term from the iterated $1\pi$-exchange and
''correct-sign'' term from the Fujita-Miyazawa three-nucleon interaction. 
One might also consider $\rho_0/2$ (where the density-gradient is maximal) as
the density relevant for estimating the spin-orbit coupling strength. At the
same time, a detailed analysis of the spin-orbit strength function $F_{\rm
  so}(\rho)$ in a chiral effective field theory approach \cite{3bodyso} leads
to a larger three-body contribution as shown by the dashed curve in Fig.\,4. With
this refined analysis the balance at $\rho_0/2 = 0.08\,$fm$^{-3}$ goes as
$-58.1\,$MeV\,fm$^5$ against $+48.3\,$MeV\,fm$^5$. This gives an impression of the
uncertainties involved.  

It has been pointed out that the strength of the short-range 
spin-orbit interaction,  $F_{\rm{so}}(\rho_0) \simeq  90\,$MeV\,fm$^5$ needed for 
(non-relativistic) nuclear structure calculations, is in perfect agreement with 
the one extracted from realistic nucleon-nucleon potentials (see Tables I and 
II in Ref.\cite{realnn}). This intimate connection between the strength of the 
spin-orbit interaction in nuclei and free $NN$ scattering is further
corroborated by the recent work of the T\"ubingen group \cite{fuchs}. 
Performing relativistic Brueckner calculations for the in-medium nucleon
scalar and vector self-energies, they find that these strong mean-fields (of
opposite sign) are almost entirely driven by the short-range spin-orbit part of
the $NN$ potential used as input. If the corresponding strength parameter is
tuned to zero both the Lorentz scalar and vector mean-fields tend to vanish
completely (see Fig.\,10 in Ref.\cite{fuchs}).  
\begin{figure}
\begin{center}
\includegraphics[scale=0.8,clip]{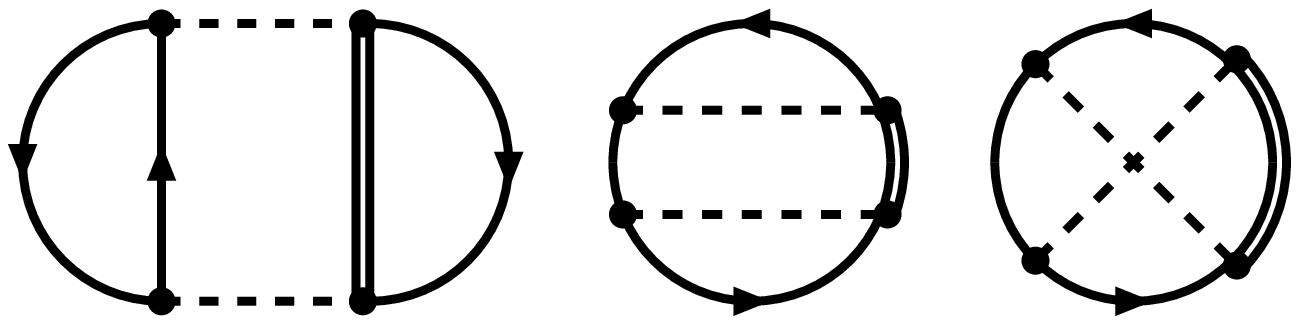}
\end{center}
\vspace{-0.2cm}
{\it Fig.\,3: Three-body Hartree and Fock diagrams of $2\pi$-exchange with
virtual $\Delta(1232)$-isobar excitation.}
\end{figure}

We thus conclude at this point that the nuclear spin-orbit interaction is 
subject to a subtle balance between three major pieces. The longer range
spin-orbit term from the iterated (2nd order) pion exchange tensor force is
canceled by the Fujita-Miyazawa type three-body contributions involving an
intermediate $\Delta$ isobar. As a consequence the 
short-range scalar and vector pieces, familiar from Dirac-Hartree phenomenology,
can account for the observed spin-orbit strength. Of course, an alternative
book-keeping would also work equally well: one could have balanced the 
scalar-vector mean field contribution against just the Hartree term from 
iterated pion exchange and built up the spin-orbit strength by the Pauli
blocking and Fock exchange pieces of in-medium second order pion exchange
together with the three-body spin-orbit force. A detailed assessment of the
interplay of all these different mechanisms becomes possible by comparing
their actions in nuclei and $\Lambda$ hypernuclei. 

\begin{figure}
\begin{center}
\includegraphics[scale=0.5,clip]{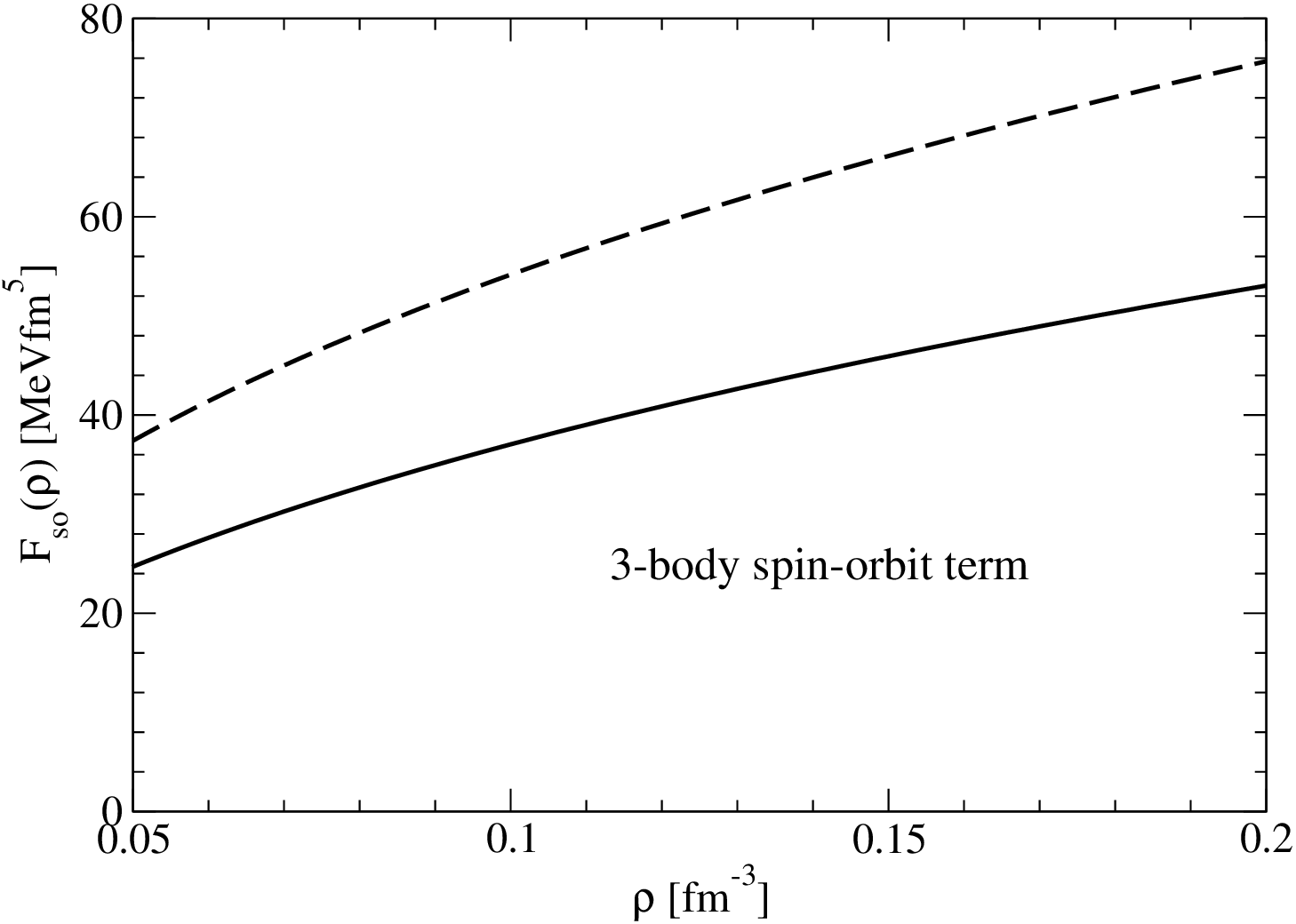}
\end{center}
\vspace{-0.2cm}
{\it Fig.\,4: Spin-orbit coupling strength generated by the $2\pi$-exchange 
three-nucleon interaction. Full curve: Fujita-Miyazawa mechanism involving
virtual $\Delta(1232)$-excitation. Dashed curve: Chiral effective field theory
approach using the low-energy constant $c_3 =-3.9\,$GeV$^{-1}$ related to the
nucleon axial polarizability \cite{3bodyso}.}
\end{figure}

\section{Spin-orbit coupling for $\Lambda$ hyperons}
The situation for $\Lambda$ hypernuclei is in a certain sense simpler since
the aspect of self-consistency for the mean field potentials is not an issue
in this case. 
A spherical mean field potential provided by the core nucleus determines
already the single-particle motion of the $\Lambda$-hyperon. Systematic 
analyses by Millener, Dover and Gal \cite{millener} have shown that the 
empirical single-particle  energies of a $\Lambda$ bound in 
hypernuclei are well described over a wide range in mass numbers by an 
attractive mean field potential of depth $U_\Lambda \simeq -28\,$MeV,
i.e. about half as strong as the one for nucleons in nuclei. On the other hand,
the $\Lambda$-nucleus spin-orbit coupling is found to be extraordinarily
weak. For example, recent precision measurements \cite{ajimura} of 
E1-transitions from $p$- to $s$-shell orbitals in $^{13}_\Lambda$C give a
$p_{3/2}-p_{1/2}$ spin-orbit splitting of only $(152\pm 65)\,$keV, to be compared
with a value of about $6\,$MeV in ordinary nuclei. 

The empirical finding that the $\Lambda$-nucleus spin-orbit coupling appears
to be negligibly small, in comparison with the strong spin-orbit interaction
of nucleons in ordinary nuclei, posed an outstanding problem in low-energy
hadronic physics. In relativistic scalar-vector mean field models a strong
tensor coupling of the $\omega$ meson to the $\Lambda$ hyperon, equal and of
opposite sign to the vector coupling, was proposed as a possible solution
\cite{tensor}, motivated by simple non-relativistic SU(6) quark model
considerations in combination with the vector meson dominance hypothesis. We can  
demonstrate that there is actually a more natural source of cancellation in the 
hypernuclear many-body problem if one takes into account the ''wrong-sign'' 
spin-orbit coupling generated by the two-pion exchange with intermediate an 
$\Sigma$ hyperon \cite{lambdapot}.       
\begin{figure}
\begin{center}
\includegraphics[scale=0.9,clip]{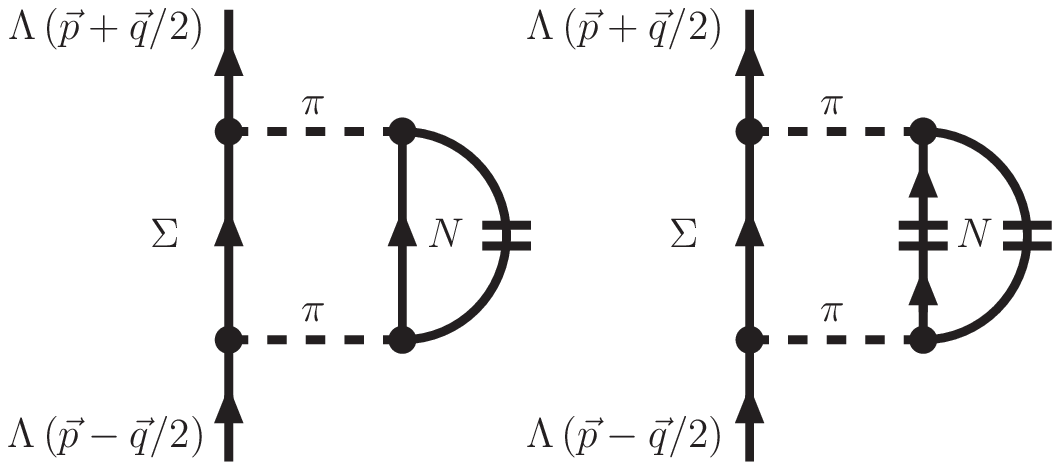}
\end{center}
\vspace{-0.2cm}
{\it Fig.\,5: Two-pion exchange processes between a $\Lambda$ hyperon and
nucleons generating a spin-orbit coupling. The horizontal double-line
symbolizes the filled Fermi sea of nucleons.}
\end{figure}

The pertinent quantity to extract the $\Lambda$-nuclear spin-orbit coupling is 
the spin-dependent part of the self-energy of a $\Lambda$ hyperon interacting
with weakly inhomogeneous nuclear  matter. Let the $\Lambda$ scatter
from initial momentum $\vec p - \vec q/2$ to final momentum $\vec p +\vec
q/2$. The spin-orbit part of the self-energy is then:    
\begin{equation} \Sigma_{\rm spin} = {i \over 2} \vec \sigma \cdot (\vec q
  \times \vec p\,) \, U_{\Lambda ls}(\rho)\,, \end{equation}
where the density-dependent spin-orbit strength  $U_{\Lambda ls}(\rho)$ is
taken in the limit of zero external momenta: $\vec p = \vec q = 0$. Its value
at nuclear matter saturation density enters as a strength parameter in the 
more familiar  spin-orbit Hamiltonian of the shell model:
\begin{equation} {\cal H}_{\Lambda ls} = U_{\Lambda ls}(\rho_0) \,\,{1 \over
2r} {df(r)\over dr}\,\,  \vec \sigma \cdot \vec L\,.
\end{equation} 
where $f(r)$ is a normalized nuclear density profile and $\vec L = \vec r
\times \vec p$ the orbital angular momentum.

The key observation \cite{lambdapot} is now that the iterated one-pion
exchange with an intermediate $\Sigma$ hyperon (see Fig.\,5) also generates a
sizeable $\Lambda$-nuclear 
spin-orbit coupling, with ''wrong" sign opposite to that from scalar and vector 
mean fields. The basic mechanism behind it is again the spin- and momentum 
dependence of the pion-baryon interaction at second order. The prefactor 
$i\, \vec \sigma \times  \vec q$ is immediately identified by rewriting the 
product of $\pi \Lambda \Sigma$-interaction vertices $\vec\sigma\cdot(\vec l- 
\vec q/2)\,\vec \sigma \cdot (\vec l + \vec q/2)$, with $\vec l$ the momentum
of the intermediate $\Sigma$ hyperon. The other factor $\vec p$ emerges from the 
non-relativistic energy denominator which includes also the small 
$\Sigma\Lambda$-mass splitting $M_\Sigma- M_\Lambda =77.5\,$MeV.

\begin{figure}
\begin{center}
\includegraphics[scale=0.6,clip]{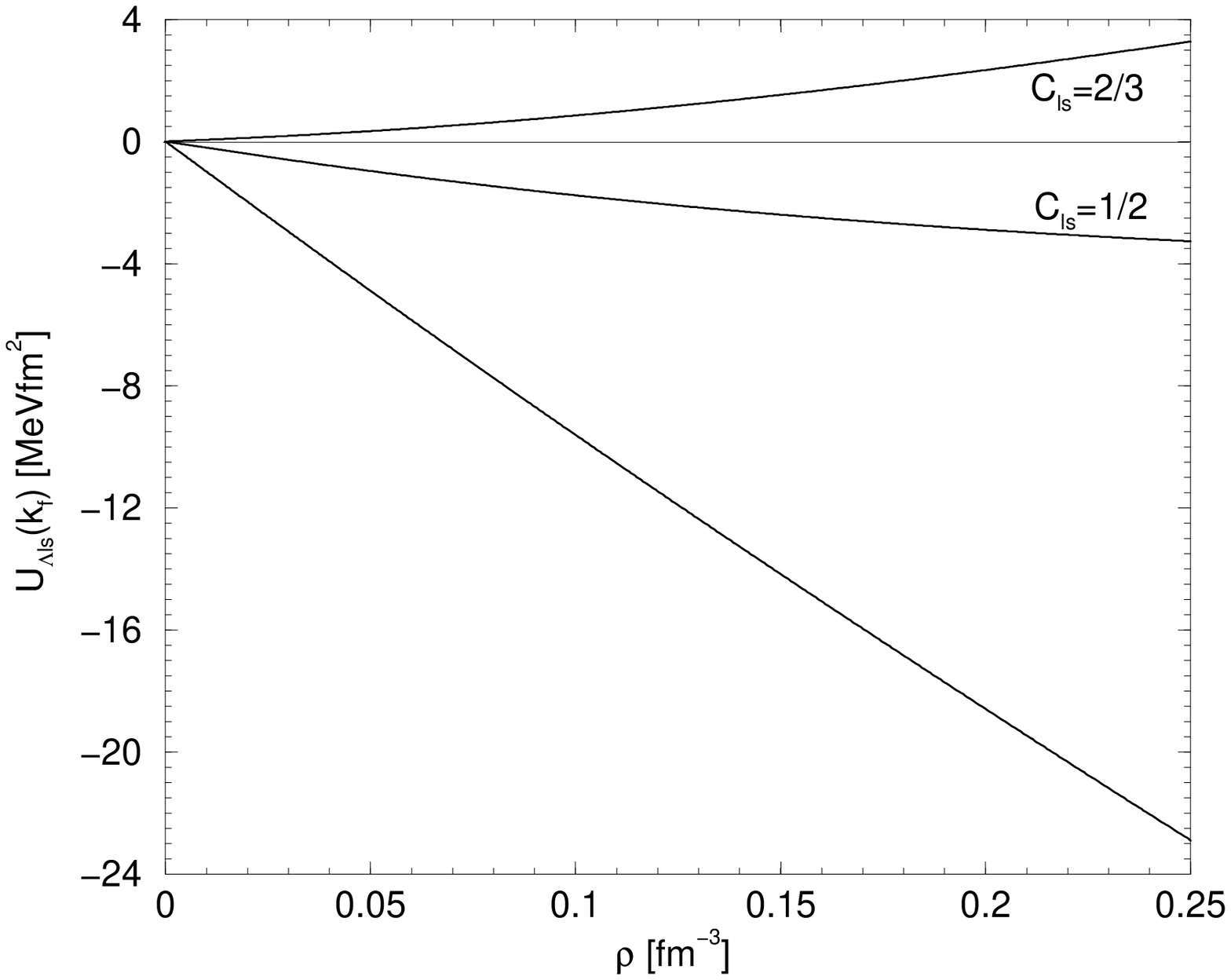}
\end{center}
\vspace{-0.2cm}
{\it Fig.\,6: The spin-orbit coupling strength $U_{\Lambda ls}(\rho)^{(2\pi\Sigma)}$ 
of a $\Lambda$ hyperon as  function of the nuclear density $\rho$. The lower 
curve shows the long-range contribution from iterated pion exchange with a
$\Sigma$ hyperon in the intermediate state. The two upper curves include in
addition the short-range  contribution,  $U_{\Lambda ls}(\rho)^{(\rm sr)} = 24.8\,C_{ls}
\,$MeV\,fm$^2\cdot \rho/\rho_0$, with $C_{ls} = 2/3$ and $1/2$ as options.}
\end{figure}
The lower curve in Fig.\,6 shows the $\Lambda$-nuclear spin-orbit coupling 
strength  $U_{\Lambda ls}(\rho)^{(2\pi\Sigma)}$ resulting from these long-range 
two-pion exchange processes as a function of the nucleon density (for
analytical expressions, see Eqs.(11-15) in Ref.\cite{lambdapot}). With a value 
of $U_{\Lambda  ls}(\rho_0) \simeq -15\,$MeV\,fm$^2$ at normal nuclear matter
density $\rho_0 =0.16\,$fm$^{-3}$ it is comparable in magnitude but of opposite 
sign with respect to the spin-orbit terms from scalar and vector mean fields. 
The upper two curves in Fig.\,6 include in addition a short-range component
of the $\Lambda$-nuclear spin-orbit coupling:    
 \begin{equation} U_{\Lambda ls}(\rho)^{(\rm sr)}=  C_{ls}{M_N^2 \over 
M_\Lambda^2} \,  U_{N ls}(\rho)^{(\rm sr)}\,. \end{equation}
which has been scaled to the one of nucleons. This coefficient $C_{ls}$ 
parameterizes the ratio of relevant coupling strengths. We have varied it
between $C_{ls}=2/3$ (following from naive quark model considerations) and $C_{ls}
= 1/2$ as suggested by in-medium QCD sum rule calculations of the Lorentz 
scalar and vector mean fields \cite{sumrule}. For the nucleonic spin-orbit 
coupling  strength we take the value $U_{N ls}(k_{f0})^{(\rm sr)} = 3\rho_0 W_0/2
\simeq 30\,$MeV\,fm$^2$ from shell model calculations. One observes from Fig.\,6
an almost complete cancellation between short-range and long-range
contributions. This balance offers a natural explanation of the empirically 
observed small spin-orbit splitting in $\Lambda$ hypernuclei. 

It is important to note that the nuclear three-body contributions which
compensated the ''wrong-sign'' spin-orbit terms (from the second order pion
exchange tensor force) does not exist in hypernuclei. The 
absence of an analogous three-body mechanism for a $\Lambda$ in the hypernucleus
becomes immediately clear by inspection of Fig.\,7. Replacing the external
nucleon by a $\Lambda$ introduces an intermediate $\Sigma$ hyperon. However,
since there is no filled Fermi sea for hyperons, a three-body force analogous
to that of Fig.\,7 does not exist in this case. Moreover,
it has recently been shown in Ref.\cite{decuplet} that the 
cancellation mechanism proposed in Ref.\cite{lambdapot} is not disturbed by 
the inclusion of analogous
$2\pi$-exchange processes with all relevant decuplet baryons ($\Delta(1232)$ and
$\Sigma^*(1385)$) in the intermediate state. These effects have alternating
signs from spin-sums and are suppressed by considerably larger mass-splittings
in the energy denominator.

The emerging picture of the nuclear and hypernuclear spin-orbit interaction is
a intriguing one. The spin-orbit interaction of nucleons is predominantly of
short range because the longer range $2\pi$-exchange components find a
mechanism of self-cancellation involving an important three-body term. 
The smallness of the $\Lambda$-nuclear
spin-orbit coupling, on the other hand, reveals the existence of  long range
$2\pi$-exchange component of the ''wrong sign'' which balances the
short-distance mechanisms. In a recent paper \cite{FKVW07} is has been
demonstrated that this scenario works indeed for actual (finite) $\Lambda$
hypernuclei.   
\begin{figure}
\begin{center}
\includegraphics[scale=1.1]{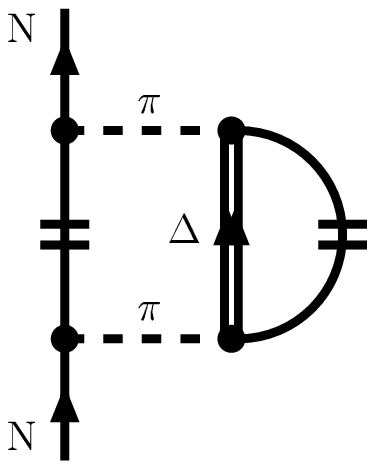}
\end{center}\vspace{-.4cm}
{\it Fig.\,7: Three-body diagram of two-pion exchange with virtual 
$\Delta(1232)$ isobar excitation. For a nucleon it generates a sizeable 
(three-body) spin-orbit force. The analogous process is not possible for a 
$\Lambda$ hyperon since there exists no filled hyperon Fermi sea.} \end{figure}

\section{Summary and conclusions}
Let us summarize our findings. It is well known that large Lorentz scalar 
and vector mean fields of opposite sign can generate (via the lower 
components of Dirac spinors) the strong spin-orbit 
coupling of nucleons in nuclei as a relativistic effect. A successful nuclear 
structure phenomenology has been developed on the basis of this mechanism. Less 
well known is the fact that two-pion exchange also produces 
sizeable spin-orbit coupling terms. The mechanism behind these is quite
different and not of relativistic origin. These spin-orbit couplings arise
directly from the spin- and momentum dependence of the pion-nucleon
interaction together with small energy denominators which enhance the effects. 
     
When working out these long range spin-orbit coupling terms (which depend 
only on well-known hadronic parameters) one finds that a ''wrong-sign''
contribution from the second order pion-exchange tensor force gets canceled
by a contribution from the three-nucleon interaction of
Fujita-Miyazawa type (mediated by $2\pi$-exchange with virtual $\Delta$ isobar
excitation). We have explicitly verified this cancellation mechanism by studying
the spin-orbit strength function $F_{so}(\rho)$ in the nuclear energy density 
functional around nuclear matter saturation density, $\rho_0=0.16\,$fm$^{-3}$.

When searching for the microscopic origin of the strong Lorentz scalar and 
vector mean fields one finds, within relativistic Brueckner calculations, that
they are mainly caused by the short-range spin-orbit part of the
$NN$ potential used as an input. The same connection is 
observed when comparing directly the spin-orbit interaction strength $3W_0/4$ 
in the Skyrme phenomenology with the one extracted from realistic
nucleon-nucleon potentials. An alternative approach to the strong Lorentz
scalar and vector mean fields in nuclear matter is provided by QCD sum rules 
which link those fields to changes of quark condensates at
finite baryon density.   

For a $\Lambda$ in a hypernucleus the balance between short-range and 
long-range components of the spin-orbit coupling is qualitatively different as
compared to ordinary nuclei. The
three-body contribution (induced by $2\pi$-exchange) is now absent. The
(scalar-vector) mean field term is canceled by a ''wrong-sign'' contribution
from  iterated pion-exchange with an intermediate $\Sigma$ hyperon. The
small mass splitting $M_\Sigma-M_\Lambda = 77.5\,$MeV is crucial
in order to make this cancellation so effective. We note in passing that a
similar cancellation mechanism is expected to be at work for the
$\Sigma$-nuclear spin-orbit interaction \cite{sigmals}. Unfortunately, the
prospects for observing this effect are poor because of the recently established
repulsive nature of the $\Sigma$-nuclear bulk potential \cite{friedgal}.

This description of spin-orbit interactions guided by chiral effective field
theory, with a short-range (mean field) 
component and long-range contributions of alternating signs from 
$2\pi$-exchange, can explain (at least qualitatively) the pronounced difference 
in strength between the spin-orbit interaction of nucleons in ordinary
nuclei and  a $\Lambda$ in hypernuclei.        

\subsection*{Acknowledgements}  We thank Avraham Gal for many stimulating
discussions.

\end{document}